\documentclass[a4paper,11pt]{article}

\usepackage[latin1]{inputenc} %définit la table de caractère latine
\usepackage[english]{babel}
\usepackage[centerlast]{caption}
\usepackage{indentfirst}
\usepackage{amsmath,amssymb,amsbsy,latexsym,textcomp}
\usepackage{bm,bbm}
\usepackage[dvips,xdvi]{graphicx}
\usepackage{pstricks}
\usepackage{subfig}

\title{\textsc{Prior knowledge and Markov parameters of linear time-invariant models}}
\author{Guillaume Mercère\thanks{G. Mercère is with Poitiers University, Laboratoire d'Automatique et d'Informatique Industrielle,
2 rue Pierre Brousse, B.P. 633, 86022 Poitiers Cedex, France. Email address: guillaume.mercere@univ-poitiers.fr}}
\date{\today}

\begin{document}

\maketitle

\section{Motivations}

The subspace-based state-space system identification techniques have been applied to different industrial applications with success for more than two decades \cite{VVG94,Bal97,AL97,Aal98,Bal98,MHA99,BAB00,FMV00}. A quick look at these contributions leads to the conclusion that these accurate results are mainly obtained with collected measurements of good quality. It is now well-known that using persistently exciting inputs (of sufficiently high order) is compulsory in order to get a reliable estimated model \cite{SS89,Lju99,Wal05}. Like any standard identification method, the subspace-based identification algorithms require the verification of specific excitation constraints in order to verify particular rank conditions and to ensure the consistency of the subspace estimates \cite{JW98,BJ00}. Unfortunately, in many practical situations, these excitation constraints are difficult to be satisfied because they may involve experimental conditions not conceivable for economical and/or safety reasons. Such a poor experimental framework often leads to a small amount of measurement samples corrupted by noise with a low signal-noise-ratio. This lack of information (due to the poor excitation of the system) should be improved by adding prior knowledge about the system into the identification procedure \cite{Rot73}. Indeed, it is common for the operator to have prior information concerning the process to be identified, \emph{e.g.}, from 
\begin{itemize}
\item its own experience,
\item the laws of physics governing the behavior of the system,
\item simple experiments such as a step response or the steady-state response to a sinusoidal input test signal.
\end{itemize}
Most of the time, this prior physical knowledge can be translated into a stability constraint, a rough value of or bounds on the DC-gain, the settling time or the dominant time constant of the system, etc. When this prior information cannot be introduced directly into the structure of the model and/or by fixing some of the parameters of the model\footnote{This is the case, \emph{e.g.}, in the subspace-based identification framework where fully-parameterized black-box models are handled.}, it is often used indirectly by formulating an optimization problem with equality or inequality constraints \cite{Rot73,Joh97}. Unfortunately, the use of regularized optimization algorithms for subspace-based identification is usually difficult because most of the subspace-based algorithms do not resort to any explicit cost functions. The incorporation of prior knowledge into the subspace-based identification is also complicated by the fact that the subspace-based state-space estimates are fully-parameterized black-box models. This property, which makes in a way the implementation of the subspace-based algorithms easier, is an important shortcoming when prior information can be used to improve the efficiency of the algorithm. The parameters of the estimated state-space matrices are indeed rarely invariants of the system. Thus, introducing prior information with a physical meaning into a fully-parameterized black-box model known up to a similarity transformation seems to be a challenging problem. In the literature, solutions have been developed in order to incorporate specific prior knowledge such as guaranteed stability \cite{Mac95,LB03}, some frequency domain constraints \cite{Hal04b} or some structural information \cite{MV08,LEL09}. Unfortunately, these first attempts cannot be used to deal with any prior knowledge. Indeed, these solutions only focus on specific constraints and are difficult to be extended as a general framework. In order to get round this limitation, it can be relevant
\begin{itemize}
\item to describe the studied subspace-based identification algorithm as an optimization-based algorithm,
\item to transform the available prior information into equality (or inequality) constraints,
\item to solve the constrained optimization problem which follows from the combination of both aforementioned steps.
\end{itemize} 
This idea has been recently put into practice in \cite{TH09,Aal10,Pal10,Aal11}. In these articles, the authors more precisely incorporate time-domain information (such as a known static gain) into some subspace-based identification algorithms. The considered subspace-based identification method used in these papers is the predictor-based subspace identification algorithm developed in \cite{PSD96} (see also \cite{Mer13} for a reminder). In \cite{TH09}, the Authors used a Bayesian framework for the incorporation of the prior information into the aforementioned CCA-type algorithm. By this way, the prior knowledge is added up as ``soft constraints'' to the information available from the experimental data. In order to avoid the non-linear optimization required by the structured weighted lower rank approximation technique \cite{SLH06} used in Step 4 of Algo. 1 in \cite{TH09}, the Authors suggested in \cite{Aal10,Aal11} translating the prior information into an equality constraint and solving the following constrained least-squares problem by using the method of weighting \cite[Chapter 22]{LH95}. Notice that the solution available in \cite{Pal10} only consists in exploiting the prior knowledge into the step dedicated to the determination of the $\mathbf{B}$ and $\mathbf{D}$ matrices.

In this brief paper, we focus on the second item listed above and aim at showing that standard prior physical knowledge like (rough) values of the DC-gain or the dominant time constant of the system can be translated as constraints on the model Markov parameters. We focus on these parameters because $(i)$ they are system invariants, $(ii)$ they play a crucial role in subspace-based model identification as illustrated by the development of the famous Kung's algorithm \cite{Kun78} or the PBSID method \cite{Chi07}. The last issue, \emph{i.e.}, the determination of a solution for specific subspace-based constrained optimization problem is devoted for a future work.

\section{Prior information and linear equality constraints}

\subsection{From the Markov parameters  to the pulse responses of the system} 

Let us consider systems, the behavior of which is assumed to be described by a standard discrete-time (DT) linear time-invariant (LTI) state-space representation 
  \begin{subequations}\label{equ:innovform}
\begin{align}
\mathbf{x}(t+1) &= \mathbf{A} \mathbf{x}(t) + \mathbf{B} \mathbf{u}(t), \\
\mathbf{y}(t) &= \mathbf{C} \mathbf{x}(t) + \mathbf{D} \mathbf{u}(t).
\end{align}
\end{subequations}
Then, the Markov parameters of an LTI model satisfying the former state-space form are defined by
\begin{equation}\label{equ:markov}
  \bm{\mathcal{M}}_{i} = \left\{
\begin{matrix}
\mathbf{D} & \text{ if } i = 0 , \\
\mathbf{C} \mathbf{A}^{i-1} \mathbf{B} & \text{ if } i > 0 .
\end{matrix}
 \right. 
\end{equation}
We say that $\left( \mathbf{A}, \mathbf{B}, \mathbf{C}, \mathbf{D} \right)$ is a realization of $\left\{ \bm{\mathcal{M}}_{i} \right\}_{i = 0}^\infty$ if the equalities in Eq.~\eqref{equ:markov} hold.

As shown, \emph{e.g.}, in \cite{Sch00}, when a unit impulse $\delta_k$ defined as
\begin{equation}
  \delta_k = 
\left\{ \begin{matrix}
1 & \text{ if } \  k = 0 \\
0 & \text{ if } \  k \neq 0
\end{matrix} \right.
\end{equation}
is applied to the input $i$ of the system and a zero signal is applied to the other inputs, we get
\begin{subequations}
\begin{align}
    \mathbf{y}(0) &= \mathbf{D}(:,i) \\
\mathbf{y}(k) &= \mathbf{C} \mathbf{A}^{k-1} \mathbf{B}(:,i), \ k = 1, 2, ... 
\end{align}
\end{subequations}
This output is generally called the pulse response of the system for an impulse at the input $i$ \cite{Kai80}. A direct consequence of this result is that the Markov parameters defined previously correspond to the pulse response coefficients of the system.

\subsection{From some specific pieces of prior information to the pulse responses of the system}

The link between the impulse response coefficients of a DT LTI system and the model Markov parameters being established, it is now time to show how different pieces of prior information can be related to the pulse response of the system quite easily and, by construction, to the Markov parameters of the system. As far as the prior knowledge is concerned, the following (non-exhaustive) list of prior information can be more precisely linked to the pulse response of the system \cite{TH09,Pal10,Aal11}:
\begin{itemize}
\item the dc-gain of the system,
\item some time constants like the rise or settling times of the process,
\item the presence of input-output zero responses for a MIMO system.
\end{itemize}

\subsubsection{DC-gain}

In practice, it is frequent that an operator using a process has prior knowledge regarding its DC-gain, at least for specific input-output couples. Furthermore, it is well-known \cite{Rot73,TH09,Aal11} that, for each input-output couple $\{i,j\}$, we have
\begin{equation}
  K_{dc}^{ij} = \sum_{k=0}^\ell \bm{\mathcal{G}}_k (i,j)
\end{equation}
where $ \bm{\mathcal{G}}_k (i,j)$, $ i \in [1, n_u]$, $j \in [1, n_y]$, $k \in [0, \ell]$, are the non zero impulse coefficients and $K_{dc}^{ij}$ the corresponding DC-gain for the specific input-output couple $\{i,j\}$. Thus, written differently, by using the previously highlighted relation between the pulse response coefficients and the Markov parameters defined in Eq. \eqref{equ:markov}, we get
\begin{equation}
  \mathbf{K}_{dc} =
\begin{bmatrix}
 K_{dc}^{11} & \cdots &  K_{dc}^{1n_u} \\
\vdots & \ddots & \vdots \\
 K_{dc}^{n_y1} & \cdots &  K_{dc}^{n_yn_u}
\end{bmatrix} =
\mathbf{D} + \sum_{k=1}^\ell  \mathbf{C} \mathbf{A}^{k-1} \mathbf{B} = \sum_{k=0}^\ell \bm{\mathcal{M}}_k 
\end{equation}
where $\ell$ must be chosen large enough to ensure that, for $i \geq \ell$, $\mathbf{C} \mathbf{A}^{i} \mathbf{B} = \mathbf{0}$. Obviously, if the user has only access to the values of $K_{dc}^{ij}$ for specific couples $\{i,j\}$, the previous relation becomes
 \begin{equation}
  K_{dc}^{ij} = \sum_{k=0}^\ell \bm{\mathcal{M}}_k (i,j)
\end{equation}
for each \emph{a priori} known couple $\{i,j\}$ where $\bm{\mathcal{M}}_k (i,j)$ stands for the element the matrix $\bm{\mathcal{M}}_k$ at the intersection of the $i$th row and $j$th column. As shown in \cite{TH09}, this idea can be easily extended to the knowledge of ratio between static gains.

\subsubsection{Time constant, damping ratio and natural frequency}

% As shown, \emph{e.g.}, in \cite{Kai80}, any SISO continuous-time transfer function can be decomposed into a combination of first order and second order terms, \emph{i.e.},
% \begin{equation}
%   H(s) = \sum_{i=1}^{n_{fo}} \frac{g_i}{s-\lambda_i} + \sum_{j=1}^{n_{so}} \frac{\alpha_j + \beta_j s}{\gamma_j s^2 + \delta_j s + \varepsilon_j}
% \end{equation}
% where $n_{fo}$ and $n_{so}$ the number of first order and second order transfer functions respectively.

In many practical cases, when the user has access to prior knowledge such as a time constant or a damping ratio, it is generally assumed that the behavior of the system can be well-approximated by a first or second order system (with or without delay and/or integrator). For instance, the following simple process models
\begin{subequations}
  \begin{align}
    G_1(s) &= \frac{K}{s}e^{-T_d s} \\
    G_2(s) &= \frac{K}{1 + \tau s}e^{-T_d s} \\
    G_3(s) &= \frac{K}{s(1 + \tau s)}e^{-T_d s} \\
    G_4(s) &= \frac{K}{(1 + \tau_1 s) (1 + \tau_2 s)}e^{-T_d s} \\
    G_5(s) &= \frac{K \omega_0^2}{\omega_0^2 + 2 \xi \omega_0 s + s^2}e^{-T_d s} \\
    G_6(s) &= \frac{K (1 + \tau_z s)}{(1 + \tau_1 s) (1 + \tau_2 s)}e^{-T_d s} \\
    G_7(s) &= \frac{K (1 + \tau_z s) \omega_0^2}{\omega_0^2 + 2 \xi
      \omega_0 s + s^2}e^{-T_d s}
  \end{align}
\end{subequations}
can be seen as the most popular ones, for instance, in the literature dedicated to easy-tuning techniques for PID controller design (see \cite{Aal93,AH05} for an interesting overview). In the following, a specific attention will be payed to the transfer functions $G_1(s)-G_5(s)$. These models are indeed generic enough to make the extension of the following developments straightforward for the other ones. Notice also that prior knowledge about the zeros of the system are quite rare in practice which reduces the interest of the transfer functions $G_6(s)$ and $G_7(s)$. Finally, because the prior knowledge of the delay can be taken into account beforehand by a standard data treatment, it will assumed hereafter that $T_d = 0$.

% Paragraph
\paragraph{First order systems}

First, let us consider the model $G_2(s)$ with $T_d = 0$ and let us assume that the time constant $\tau$ is known. Then, by using the standard $z-$transform with a zero-order hold, the discrete-time (DT) transfer function counterpart of $G_2(s)$ satisfies \cite{Aal11}
\begin{equation}
  G_2(z) = K \frac{1 - e^{- \frac{T_s}{\tau}}}{z - e^{- \frac{T_s}{\tau}}}
\end{equation}
where $T_s$ stands for the sampling period of the sampler. It is straightforward to show that this system satisfies the following DT state space representation
\begin{subequations}
  \begin{align}
    x(t+1) &= e^{- \frac{T_s}{\tau}} x(t) + K \left( 1 - e^{- \frac{T_s}{\tau}} \right) u(k) \\
    y(t) &= x(t) .
  \end{align}
\end{subequations}
From this state-space form, it is easy to see that the corresponding Markov parameters verify
\begin{align}
  \mathcal{M}_0 &= 0 & \mathcal{M}_i &= \left( e^{- \frac{T_s}{\tau}} \right)^{i-1} K \left( 1 -e^{- \frac{T_s}{\tau}} \right), \ i \geq 1 .
\end{align}
Of course, these relations can be used directly but, they can also be rewritten as a recurrent equation
\begin{align}\label{equ:markparamforder} 
  \mathcal{M}_0 &= 0 & \mathcal{M}_1 &= K \left( 1 -e^{- \frac{T_s}{\tau}} \right) & \mathcal{M}_i &= e^{- \frac{T_s}{\tau}} \mathcal{M}_{i-1} , \ i \geq 2 .
\end{align}
The relations $\mathcal{M}_i = e^{- \frac{T_s}{\tau}} \mathcal{M}_{i-1} , \ i \geq 2,$ can indeed be used without knowing the gain $K$.

Now, let us focus on the integrator transfer function $G_1(s)$. By using the same approach of the one applied to $G_2(s)$, we get
\begin{equation}
  G_1(z) =\frac{K T_s}{z - 1}
\end{equation}
or, in the state-space,
\begin{subequations}
  \begin{align}
  x(t+1) &= x(k) + K T_s u(t) \\
y(k) &= x(k)
  \end{align}
\end{subequations}
which straightforwardly leads to
\begin{align}\label{equ:markparamint}
  \mathcal{M}_0 &= 0 & \mathcal{M}_i &= K T_s = cst, \ i \geq 1 .
\end{align}
Again, without knowing $K$, it is possible to use the fact that, for $i \geq 1$, the Markov parameters are constant and all equal.

These equalities (see Eq. \eqref{equ:markparamforder} and \eqref{equ:markparamint}) can be easily translated into matrix equality constraints if, for a MIMO system, the available prior information concerns several input-output couples.

% Paragraph
\paragraph{Second order systems}

The transfer functions $G_3(s)$, $G_4(s)$ and $G_5(s)$ can be studied in one shot. Indeed, by assuming (again) that the sampler contains a zero-order hold, it is quite easy to show that, for $j = {3,4,5}$, \cite{Oga09}
\begin{equation}
  G_j(z) = \frac{\beta_1 z + \beta_0}{z^2 + \alpha_1 z + \alpha_0}
\end{equation}
where
\begin{subequations}
  \begin{align}
\left.
    \begin{matrix}
      \beta_1 = K \left( T_s - \tau \left( 1 - e^{- \frac{T_s}{\tau}} \right) \right) \\
      \beta_0 = K \left( \tau \left( 1 - e^{- \frac{T_s}{\tau}} \right) - T_s e^{- \frac{T_s}{\tau}} \right) \\
      \alpha_1 = -1 - e^{- \frac{T_s}{\tau}} \\
      \alpha_0 = e^{- \frac{T_s}{\tau}} \\
    \end{matrix} \right\} & \text{ for } G_3(s) \\
\left.
    \begin{matrix}
      \beta_1 = K \frac{\tau_1 \left( 1 - e^{- \frac{T_s}{\tau_1}} \right) - \tau_2 \left( 1 - e^{- \frac{T_s}{\tau_2}} \right)}{\tau_1 - \tau_2} \\
      \beta_0 = K e^{- \frac{T_s}{\tau_1}} e^{- \frac{T_s}{\tau_2}} - K \frac{\tau_1 e^{- \frac{T_s}{\tau_2}} - \tau_2 e^{- \frac{T_s}{\tau_1}}}{\tau_1 - \tau_2} \\
      \alpha_1 = - e^{- \frac{T_s}{\tau_1}} - e^{- \frac{T_s}{\tau_2}}\\
      \alpha_0 = e^{- \frac{T_s}{\tau_1}} e^{- \frac{T_s}{\tau_2}} \\
    \end{matrix} \right\} & \text{ for } G_4(s) \\
\left.
    \begin{matrix}
      \beta_1 = 1 - e^{- \xi \omega_0 T_s} \left( \cos(\omega_p T_s) + \frac{\xi \sin(\omega_p T_s)}{\sqrt{1 - \xi^2}} \right) \\
      \beta_0 = e^{- 2 \xi \omega_0 T_s} + e^{- \xi \omega_0 T_s} \left( \frac{\sin(\omega_p T_s)}{\sqrt{1 - \xi^2}} - \cos(\omega_p T_s) \right) \\
      \alpha_1 = - 2 e^{- \xi \omega_0 T_s} \cos(\omega_p T_s) \\
      \alpha_0 = e^{- 2 \xi \omega_0 T_s} \\
    \end{matrix} \right\} & \text{ for } G_5(s)
  \end{align}
\end{subequations}
where $\omega_p = \omega_0 \sqrt{1 - \xi^2}$. By using a standard controller state-space form \cite{Kai80}, we obtain
\begin{subequations}
  \begin{align}
    x(t+1) &= 
\begin{bmatrix}
- \alpha_1 & - \alpha_0 \\
1 & 0
\end{bmatrix} x(t) + 
\begin{bmatrix}
1 \\
0
\end{bmatrix} u(t) \\
y(t) &= 
\begin{bmatrix}
\beta_1 & \beta_0 
\end{bmatrix} x(t)
  \end{align}
\end{subequations}
from which the following Markov parameters can be extracted
\begin{align}
  \mathcal{M}_0 &= 0 & {\mathcal{M}}_i &= \begin{bmatrix}
\beta_1 & \beta_0 
\end{bmatrix}\begin{bmatrix}
- \alpha_1 & - \alpha_0 \\
1 & 0
\end{bmatrix}^{i-1} \begin{bmatrix}
1 \\
0
\end{bmatrix}, \ i \geq 1 .
\end{align}
By looking closer at these parameters, it is interesting to notice that
\begin{subequations}
  \begin{align}
    \mathcal{M}_0 &= 0 & {\mathcal{M}}_1 &= \beta_1 \\
    {\mathcal{M}}_2 &= \beta_0 - \alpha_1 \beta_1 &
    {\mathcal{M}}_i &= - \alpha_1 {\mathcal{M}}_{i-1} -
    \alpha_0 {\mathcal{M}}_{i-2}, \ i \geq 3 .
  \end{align}
\end{subequations}
Notice also that the equality ${\mathcal{M}}_i = - \alpha_1 {\mathcal{M}}_{i-1} - \alpha_0 {\mathcal{M}}_{i-2}, \ i \geq 3$, can be used as a constraint only handling the parameters $\alpha_1$ and $\alpha_0$. This way of using this equality is especially interested for the models $G_3(s)$ and $G_4(s)$ because these parameters are only dependent on the time constants of the system.

% SUBSUBSECTION %
\subsubsection{Zero transfer function}

In practice, a specific input may not affect specific outputs of the system. This feature results in a zero transfer function for this input-output couple. Unfortunately, when real noisy data are used, most of the identification algorithms lead to non-zero transfer functions for these input-output zero responses when no constraints is added. Again, the standard subspace-based algorithms do not depart from this rule. It is not trivial to ensure that, for this input-output channel, the zero response is kept into the parameterization \cite{Qin06}. On the contrary, it is really easy to relate a zero transfer function to the Markov parameters. Indeed, if it is \emph{a priori} known that, for the input-output couple $\{i,j\}$, $G_{ij}(s) = 0$, then, obviously, all the coefficients of the corresponding impulse response are equal to zeros, \emph{i.e},
\begin{equation}
  \bm{\mathcal{M}}_k(i,j) \ \forall \ k \geq 0 .
\end{equation}
% As shown hereafter, such prior known constraints on the components of the Markov parameters can be translated into matrix equalities.

\section{Conclusion}\label{para:concl}

In many practical cases, the engineer has access to prior knowledge like rough values of the DC-gain or the main time constant of the system. In order to improve the accuracy of subspace-based identification techniques using the model Markov parameters, we derive in this short paper the direct links between these impulse response coefficients and this prior information. The next step will consist in introducing this prior knowledge explicitly in Kung's algorithm thank to dedicated equality and equality constraints. This issue is devoted to a future work.

\bibliographystyle{plain}
% \bibliography{/home/gmercere/Documents/Research/ScientificProductions/Bibliography}

\end{document}